\documentclass{sig-alternate-10pt}

\usepackage[hyphens]{url}
\usepackage{xspace}
\usepackage{graphicx}
\usepackage{subfig}
\usepackage[binary-units=true]{siunitx}
\usepackage{xcolor}
\usepackage{paralist}

\newcommand{\nettovec}{Net2Vec\xspace}

\begin{document}

\date{}

\title{\nettovec: Deep Learning for the Network}

\author{
{\rm Roberto Gonzalez, Filipe Manco, Alberto Garcia-Duran, Jose Mendes, Felipe Huici, Saverio Niccolini, Mathias Niepert}\\
NEC Labs Europe\\
{first.last}@neclab.eu
} 

\maketitle


\begin{abstract}
We present \nettovec, a flexible high-performance platform that allows
the execution of deep learning algorithms in the communication
network. \nettovec is able to capture data from the network at more
than 60Gbps, transform it into meaningful tuples and apply predictions
over the tuples in real time. This platform can be used for different
purposes ranging from traffic classification to network performance
analysis.

Finally, we showcase the use of \nettovec by implementing and testing
a solution able to profile network users at line rate using traces
coming from a real network. We show that the use of deep learning for
this case outperforms the baseline method both in terms of accuracy
and performance.

\end{abstract}

\section{Introduction}
\label{sec::introduction}
Deep learning approaches have been successfully applied in a wide
range of areas such as computer vision, natural language processing,
and resource management, to name a few. The success of deep learning
is largely explained by its ability to learn vector representations of
the objects under consideration (images, words, system states). These
representations are highly suitable for the downstream classification
and regression problems.

As a result, a number of deep learning frameworks have risen in the
past few years (e.g., \textsc{Torch} and \textsc{TensorFlow}) that
enable data scientists to prototype, train, and evaluate neural
network models. Further, these frameworks are optimized for batch
processing, where large data sets are moved to and processed on GPUs.

However, to the best of our knowledge, there is no platform that
targets, and is optimized for, the development and deployment of deep
learning methods in \emph{data communication networks}, with their
high throughput and low latency requirements. While there are some
platforms for stream processing such as
\textsc{Flink}~\cite{carbone2015apache}, these have overheads due to
fault tolerance considerations and/or the enforcement of in-sequence
processing of the data streams. For most of the deep learning
algorithms, however, in-sequence processing is not a strict
requirement. Instead, the sequences only have to be processed in an
order that approximates the original one.  For instance, most
representation learning methods~\cite{mikolov2013distributed} are
agnostic to violations of the original order if these violations are
local (e.g., permutations of two neighboring words). There are no
implementations of deep learning methods for existing stream
processing platforms.

A platform optimized for network analytics would be highly beneficial
as many of the typical problems could be addressed with deep learning
methods. Some of these problems are CDNs/content popularity
prediction, anomaly detection (e.g., DoS attacks, port scans, etc.),
online advertisement, advanced traffic classification, and fault
prediction, to name a few.

To plug this gap, we introduce \nettovec, a proof-of-concept
implementation of a high performance platform that brings deep
learning to the network. \nettovec can work both on live traffic,
using a high performance traffic capture module (at least 60Gb/s in
our tests), or off-line, using modules that retrieve information from
data or log files. In the case of live traffic, \nettovec provides
modules that transform incoming network packets into sequences of
tuples; deep learning models are then applied to the tuple sequences
to learn representations for downstream analytics tasks. \nettovec
also supports the acceleration of ML algorithms through the use of
GPGPUs.

To showcase \nettovec's performance and applicability, we use it to
implement a user profiling use case. We show that it outperforms a
baseline method both in terms of efficiency and accuracy of results.

\section{Design and Implementation}
\label{sec::design}

\begin{figure*}[t!]
\centering
\includegraphics[width = 0.8\textwidth]{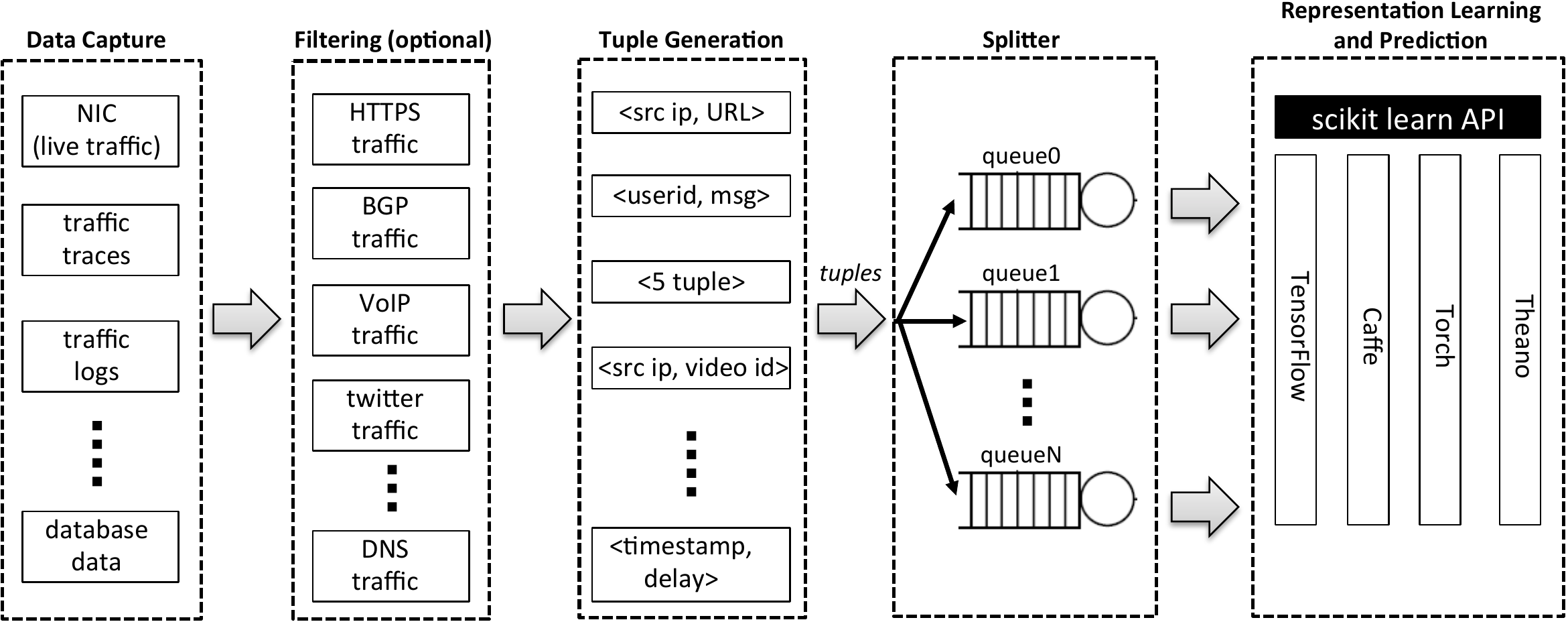}
\caption{\label{fig-arch} \nettovec architecture, showing components
  as dotted line rectangles and modules as solid rectangles within
  them. Network traffic data are captured from network interfaces or
  other sources, optionally filtered, and sequences of tuples
  generated from them (first three components). A splitter takes a
  user-provided key to split the tuples into separate tuple streams
  that are then analyzed by representation learning algorithms.}
\end{figure*}


The main objective of \nettovec is to provide data scientists with a
framework to develop and deploy machine learning methods in
networks. Network traffic consists of high-speed packet sequences (or
data derived from them such as log files) carrying high-level information
such as sequences of text, images, speech and so on. State of the art
(deep) learning algorithms have shown remarkable performance for
analytics tasks on such sequences of objects, and so \nettovec aims to
bring these state of the art ML methods to bear on communication
networks. 


At a high level, \nettovec is in charge of capturing packet data,
filtering it, constructing tuples from it, and feeding those tuples
to the machine learning algorithms in charge of the analysis. More
specifically, \nettovec's architecture consists of a set of five
\emph{components}, each containing pluggable \emph{modules} (see
Figure~\ref{fig-arch}):

%
 \noindent\textbf{Data Capture}: This component is in charge of capturing
   network data from a number of sources including network interfaces,
   traffic trace files or log files. Our proof-of-concept evaluation
   implements a Netmap~\cite{netmap} based module that captures live
   traffic from network interfaces.

  \noindent\textbf{Filtering}: A set of modules responsible for filtering out
    uninteresting traffic. This component is optional: depending on
    the data capture module, it may be that all traffic is
    relevant. In our evaluation we implement a module that filters for
    HTTP traffic.

  \noindent\textbf{Tuple Generation}: This component transforms the
    incoming packet data stream into a set of tuples suitable for the
    representation learning component. For example, in the user
    profiling use case described further in this paper, the tuples are
    of the form \texttt{<src ip, hostname>}.

\noindent\textbf{Splitter}: This component takes, as input from users, a
    subset of attributes and uses it to split the incoming tuple
    stream into separate sequences of tuples. Analogous to the
    relational database terminology, we refer to these subsets of
    attributes as \emph{keys}. The individual sequences become the
    input to the representation learning methods.

\noindent\textbf{Representation Learning and Prediction}: This component supports any
    algorithm that complies with the basic methods of the scikit-learn 
    API
    . In addition to the
    standard ML algorithms, most deep learning frameworks have
    wrappers that expose these methods. The input to the ML models are
    fixed-size (possibly padded) sequences of tuples, one per key
    value.

Users of the platform can implement data analysis use cases by
choosing modules for the different components described above, along
with a specification of a key to be used by the splitter. Extending
the platform's functionality can be done by developing additional
modules. In the rest of this section we give a more detailed
explanation of the various components, as well as a performance
evaluation of the data capture, filtering and tuple generation components.

\subsection{Capture, Filtering and Tuple Generation Components}
\nettovec targets network data, and as such, the platform needs to be
able to cope, in real-time, with the sort of traffic rates typical of
operator networks. Specifically, the aim is to show that \nettovec is
able to capture traffic from multiple interfaces, perform filtering on
it in order to extract the traffic relevant to specific use cases,
and transform the result into the tuples needed by the representation
learning algorithms.

To demonstrate this, we implement a \nettovec high-performance packet
capture module based on the Netmap framework~\cite{netmap}. An
additional module then applies a filter (in our case \texttt{dst port
  80}), and a third one parses the resulting packets to generate a
tuple (a two-element tuple consisting of \texttt{<src ip, hostname>}.

To test the performance of these modules, we use an x86 server with an
Intel Xeon E5-1650 v2 3.5 GHz CPU (6 cores), 16GB of RAM and 3 Intel
X520-T2 10Gb/s dual port cards (a total of 60 Gb/s). For packet
generation we use a separate server also with 3 Intel X520-T2 cards,
and connect each of the 6 ports to those on the server running
\nettovec with direct cables. For packet generation we use the
\texttt{pkt-gen} netmap application. In addition to the
filtering/tuple described above, we measure baseline capture
performance by asking the module to receive packets, calculate a rate,
and discard them.


\begin{figure}[t!]
\centering
\includegraphics[width = 0.48\textwidth]{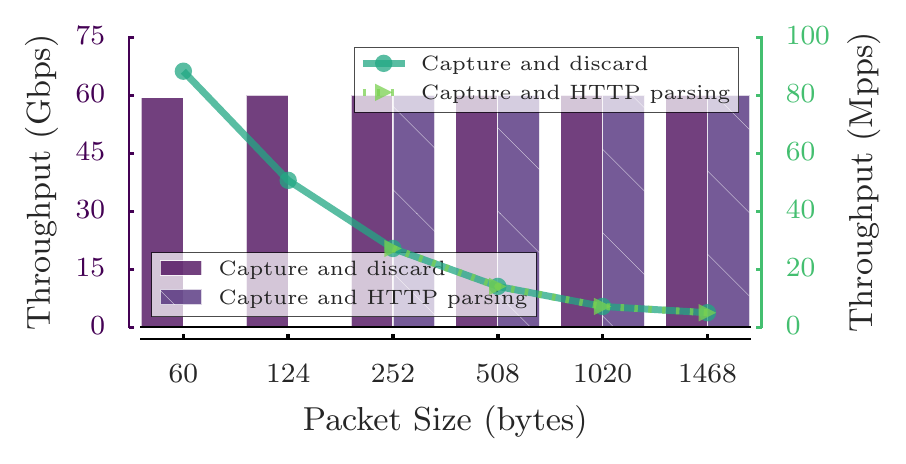}
\caption{\label{fig-capture} Performance of the \nettovec high
  performance packet capture module. The bars correspond to the left
  y-axis (throughput in Gb/s), the curves to the right one (Mp/s)}
\end{figure}

The results are shown in Figure~\ref{fig-capture}. The baseline
experiment (``Capture and discard'') shows that the module is able to
achieve line rate (60 Gb/s) for all packet sizes (except for a slight
drop for minimum-sized ones). The full experiment (``Capture and HTTP
parsing'') yields similar results: line rate for all packet sizes
(note: in this case there are no results for small packets sizes due
to headers and the HTTP payload). In short, \nettovec is able to
handle large traffic rates while generating the tuples needed by the
machine learning algorithms.

\subsection{Splitter Component}
The splitting component takes the user provided \emph{key} and splits
the incoming global stream of tuples into key-specific streams. Each
stream is associated with a queue of size $n$. Once a queue is filled,
the content of the queue (which is a sequence of tuples) is sent to
the ML component to update its parameters and/or to perform a
prediction. The oldest tuple in the queue is removed. The splitting of
the traffic with respect to some key is crucial in several use cases
such as user profiling and can also serve load balancing purposes.

In the user profiling use case we will describe later, the \emph{key} consists
of the source IP address which we use to uniquely identify individual
users. The queue stores sequences of hostnames that are used to train
a deep unsupervised embedding model.

\subsection{Representation Learning and Prediction Component}

The component accepts a machine learning model implemented as a Python class with the standard scikit-learn API methods such as \texttt{predict}, \texttt{fit}, etc. \nettovec applies the model to the incoming stream of tuples so as to learn a representation for the problem specific objects (images, hostnames, etc.) and to perform prediction for the given task. We focus specifically on neural network based machine learning methods and the corresponding frameworks such as  \textsc{Torch} and \textsc{TensorFlow}. 
These parameters of the neural models are trained  using highly efficient gradient
descent type algorithms~\cite{bottou2010large} which can be
parallelized~\cite{recht2011hogwild} and implemented for GPUs using
platforms such as CUDA~\cite{raina2009large,jia2014caffe}.
Since the ML models have to be trained and queried
at network speed, we use the cudamat
library~\cite{cudamat} and similar libraries for GPU processing.


Machine learning models implicitly perform some form of
representation learning. The input representations of objects such as
words, hostnames, images, and so on, are mapped to a vector
representation (an embedding) that is amenable to downstream
classification or regression problems. Representation learning is
especially useful in environments with an abundance of data but with
few labeled examples. Unsupervised representation learning based on
neural networks has been used in many different
contexts~\cite{lee2009convolutional,raina2009large,mikolov2013distributed,radford2015unsupervised}. \nettovec continuously feeds the incoming tuple sequences to the ML models by calling the method \texttt{update} (not included in the scikit-learn API) which performs a parameter update. These updates refine the learned representations and are able to capture concept drift.


Once the representations (embeddings) are learned, the system can
perform the use case specific prediction task. In many use cases,
labeled data only exists for a small set of tuples or sequences of
tuples. The problem is then to predict the labels for those (sequences
of) tuples that do not have labels. Given the representations of
tuples in an Euclidean space (the embedding space), there are generally
two ways to incorporate labeled data.
One method simply performs a form of $k$-nearest neighbor search and
assigns the categories of the most similar tuples to the unlabeled
tuple. The alternative strategy trains a more sophisticated ML
model using the learned embeddings as input representations and existing labels as supervision signal. The \texttt{predict} (and, if probabilities are required, also \texttt{predict\_probab}) methods are called by \nettovec to perform the prediction.

\section{Use Case: User Profiling}
\label{sec::eval}

To evaluate the implemented \nettovec platform, we assess its
performance on a user profiling use case, a typical
network analytics task that enables network operators to participate
in the lucrative online advertising ecosystem by serving targeted ads to users.
In principle, network operators have an advantage over
other user profiling players since they have access to their users'
navigation history; this is even true if most of the navigation
history is based on HTTPS requests~\cite{gonzalez2016user}.  The size
of the data carried over the network and the privacy implications of
storing user-specific data, however, make the exploitation of network
data challenging to operators.



To address this, we use \nettovec to efficiently profile users based
on real network data while maintaining only one ML model and no
long-term user-specific data.

\begin{figure}[t!]
\centering
\includegraphics[width = 0.34\textwidth]{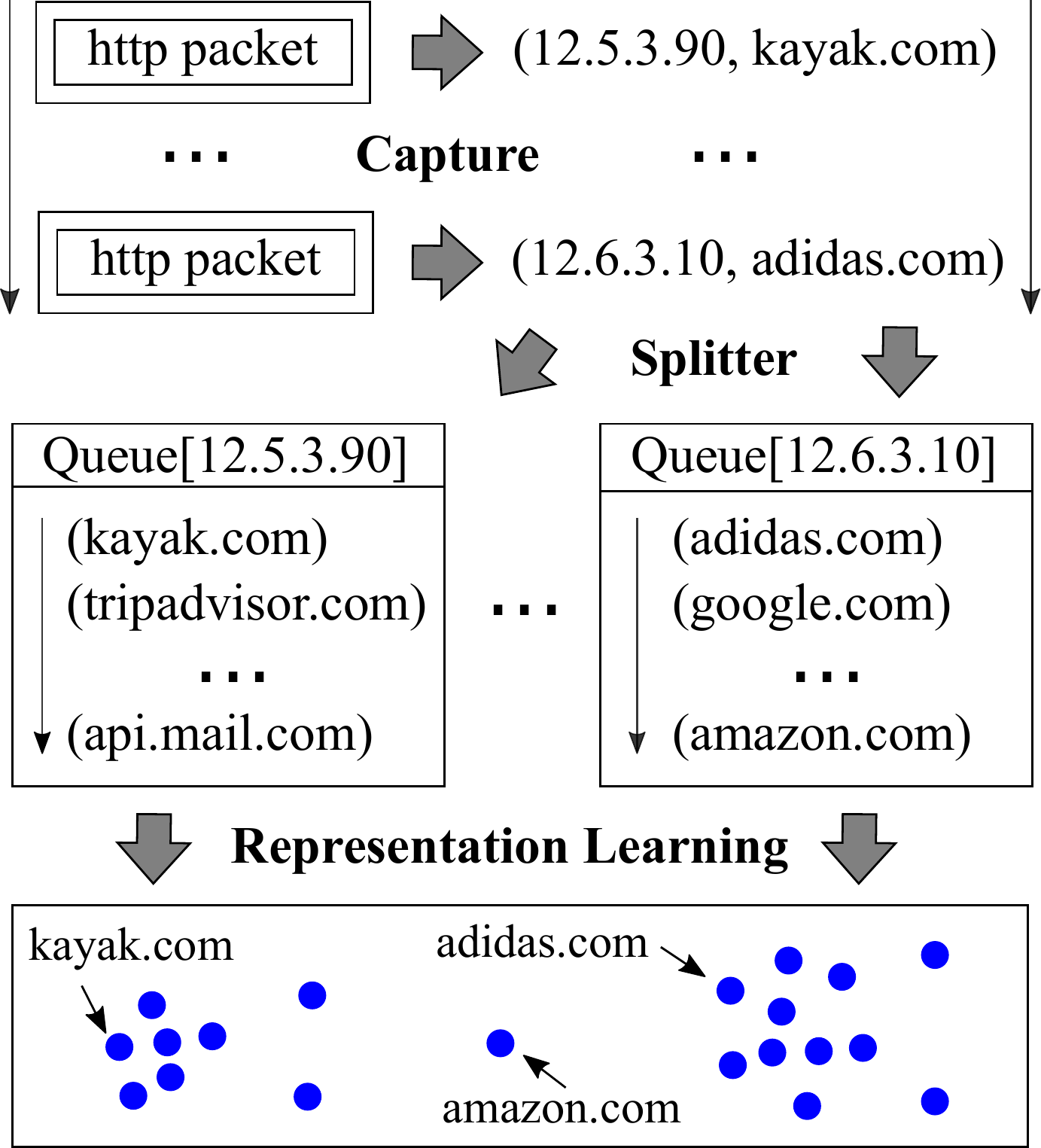}
\caption{\label{fig-net2vec} Illustration of the proposed framework using the user profiling example. \nettovec transforms sequences of packets to sequences of tuples which are split into a set of queues. The sequences of tuples maintained in each queue are fed to ML models.}
\end{figure}

\subsection{Problem Description}
For this use case we assume a network operator capturing customer
HTTP(S) traffic. The HTTP(S) requests are captured by \nettovec and
mapped into tuples of the form (\texttt{src ip, hostname}). Each tuple
represents a hostname visited by a user in the network.

In addition, the operator maintains a set of product-related
categories $\mathbf{C}$ and a subset of the hostnames is associated
with some categories from $\mathbf{C}$. Note that the number of
hostnames for which categories are known is small compared to all
known hostnames since many hostnames do not correspond to webpages but
to CDNs, trackers, and mobile application APIs. The objective is now
to assign, in real-time, a given user (here: IP address) to a set of
product-related categories based on her current URL request sequence.

The standard approach would be to follow a strategy similar to the one used
by online trackers, where all hostnames visited are stored and used to
assign categories to users. This approach, however, has several
drawbacks. First, it does not scale well since the amount of data per
user grows continually. Second, storing the browsing history of
users raises important difficult issues and is even illegal in several
countries.

To overcome these problems, we use \nettovec to train a neural network
model that learns the behavior of users. It
is then deployed to predict, given a request sequence of a user, the
product-related categories the user is likely to be interested in.
The neural network learns a representation of the hostnames from a
large number of request sequences. For each input sequence, the model
is trained to reconstruct one-hot encoding of the hostname from the rest of the
sequence. This ML model is similar to the word2vec model for learning word embeddings~\cite{mikolov2013distributed}. Figure~\ref{fig-net2vec} depicts the \nettovec system for user profiling.

\begin{figure}[t!]
\centering
\includegraphics[width = 0.39\textwidth]{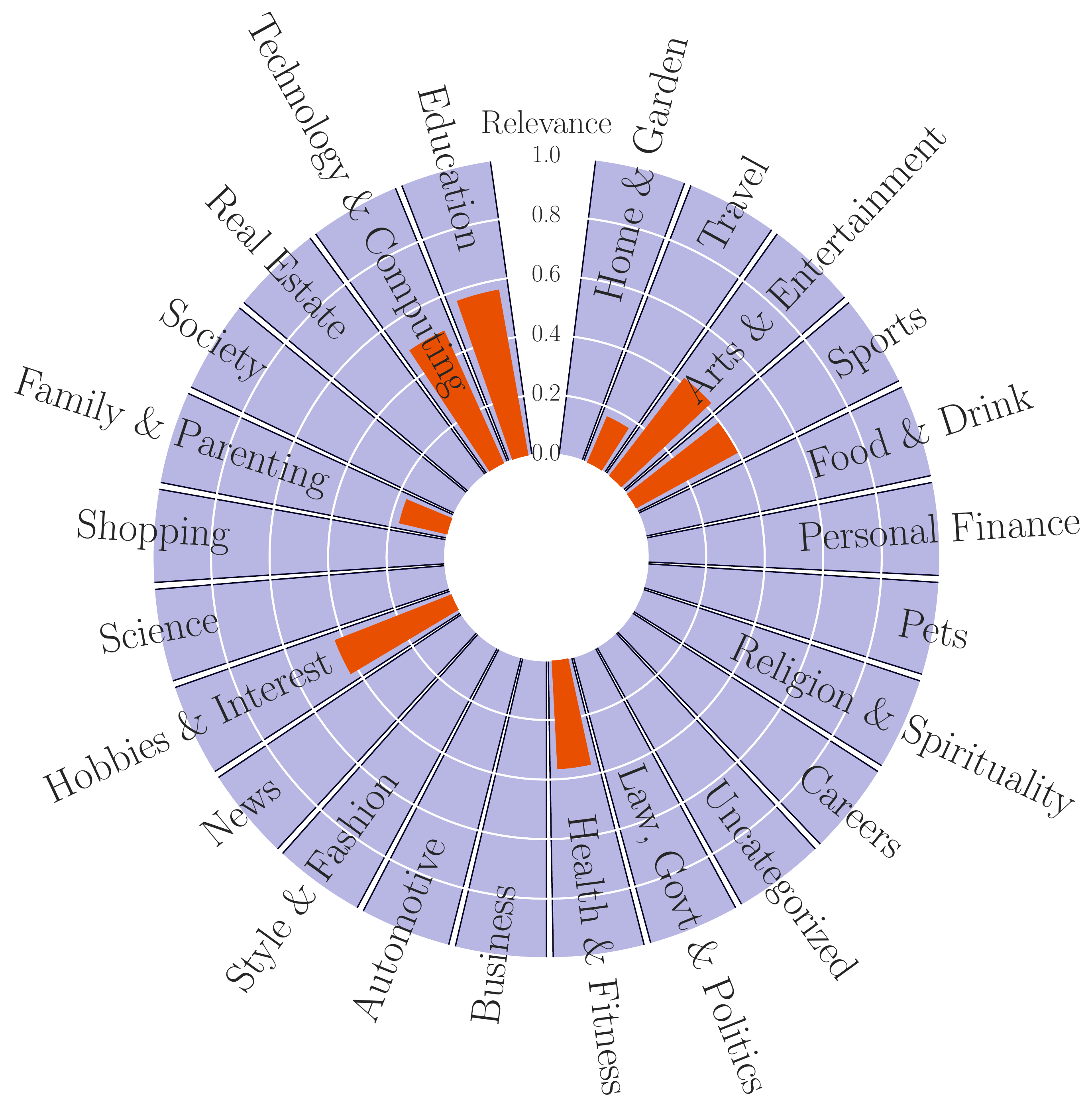}
\caption{\label{fig_profile} Profile generated for a video game persona.}
\end{figure}

\subsection{Experiments \& Results}

For the experiments we used the first level IAB
categories\footnote{The Internet Advertising Bureau (IAB) categories
  are considered to be the standard for the online advertising industry.}.
We bootstrapped the system by assigning IAB categories to hostnames
using a text classification algorithm.

We generated artificial \textit{personas} with specific interests
according to the work by Carrascosa et al. \cite{carrascosa2015always} and evaluated
the categorization provided by \nettovec for these personas.  Figure
\ref{fig_profile} shows a single profile returned by \nettovec for a
persona interested in video games. The profile indicates that the user is
mostly interested in \textit{Technology \& Computing}, \textit{Hobbies
  \& Interests} and \textit{Arts \& Entertainment}, topics directly
related with video games. In addition, the system also
predicts interests in \textit{Education}, \textit{Sports} and
\textit{Health \& Fitness}. A manual inspection of the request traces
reveals that most users in the network who visited webpages related
to video games also visited webpages related to education and
sports. The instance of the \nettovec system, therefore, is able to
assign categories to users based on the general user behavior observed
in one specific network. Similar results have been obtained for other
personas.


\begin{figure}[t!]
\centering
\includegraphics[width = 0.48\textwidth]{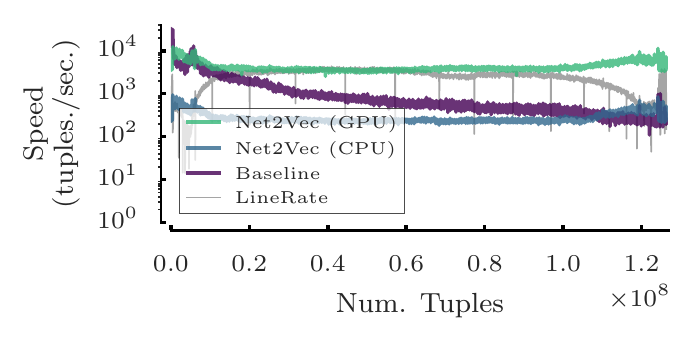}
\caption{\label{fig_perfonmand} Number of requests served from the baseline and the deep learning model for \nettovec.}
\end{figure}

We evaluated the performance of \nettovec (running only on a CPU and
accelerated by a GPU) by comparing it to a standard baseline method
using real traces from a mobile network operator. We used an x86 server with an Intel Xeon E5-2637 v4 3.5Ghz CPU, 128GB of RAM and a GeForce GTX Titan X (900 Series) GPU with a processing power of 6Tflops. We assigned
product-related categories to 200k hostnames using the text
classification algorithm, and trained \nettovec on the request
sequences of one day of traffic, predicting the categories for users
active during the following day. The trace for the second day is
composed of about 125M HTTP(S) requests initiated by more than 40k
users. Figure \ref{fig_perfonmand} depicts the number of tuples
\nettovec was able to process per second. At the beginning, the
standard baseline that simply collects categories of visited hostnames
is slightly more efficient than \nettovec executed on a GPU. However,
the baseline method's performance decreases over time since it has to
maintain all of the encountered categories per user. The performance
of \nettovec remains constant over time and is able to analyze all the
traffic of the 10Gb pipe the data came from. This highlights an
intrinsic advantage of neural network based methods: prediction time
remains constant.


\begin{figure}
\centering
\includegraphics[width = 0.48\textwidth]{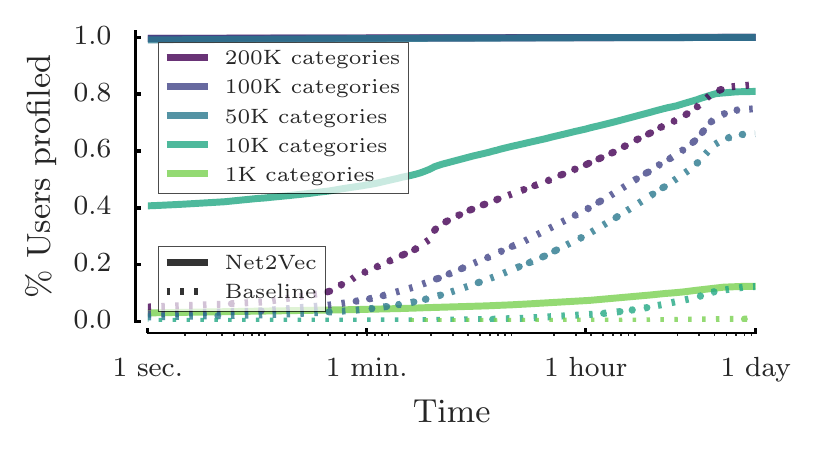}
\caption{\label{fig_timeToProfile} Time needed to generate a profile with Net2Vec versus a baseline scenario for different number of categorized hostnames.}
\end{figure}

A crucial advantage of the \nettovec deep learning model for user
profiling is that it is able to generate a profile for a user even if
none of the hostnames visited by the user have assigned categories. In
Figure \ref{fig_timeToProfile} we present the \% of users profiled
after a given amount of time using both \nettovec and the baseline
approach for different number of categorized hostnames. When we have
categories for at least 50K hostnames, the \nettovec system is able to
generate a profile for all the users since the first HTTP
request. Moreover, even when faced with only 10K categorized hostnames, the
system needs less than 1 hour of traffic to generate the profile of
70\% of the users.  The baseline approach performs much worse, having
to process more than 1 hour of traffic before being able to generate a profile for most
of the users. Moreover, even with all the traffic of the day, the
baseline approach cannot generate a profile for more than 20\% of the
users (even when 200K hostnames are categorized).

\begin{figure}
\centering
\includegraphics[width = 0.48\textwidth]{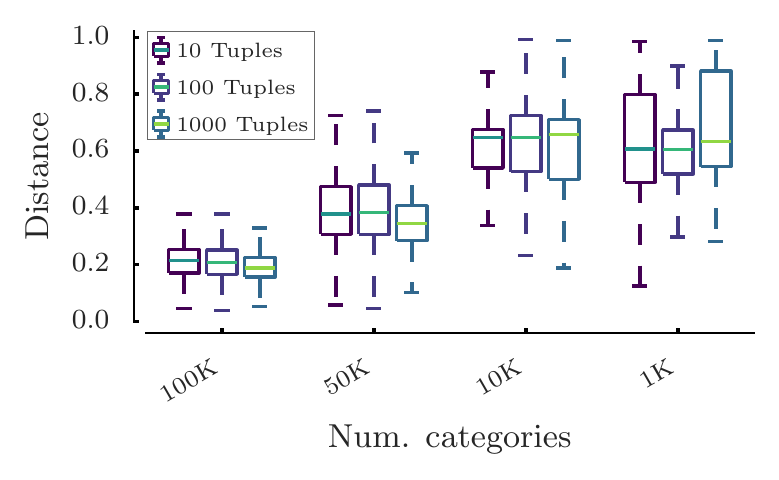}
\caption{\label{fig_RelAccuarcy} Profile similarity for different number of hostnames categorized.}
\end{figure}

We also evaluate the robustness of the approaches to changes in the
number of hostnames for which categories are known before the user
profiling starts. This robustness is important since, increasingly,
hostnames are not amenable to a content-based assignment of categories
(e.g., due to more mobile app traffic and less web browser
traffic). We computed the distance between the profiles generated with
100k, 50k, 10k, and 1k previously categorized hostnames with the
profiles generated using 200k previously categorized hostnames. We do
this comparison for every user after $10$, $100$, and $1000$ observed
HTTP(S) requests. One can observe that a small number of requests is
sufficient to generate high quality profiles. The number of previously
categorized hostnames has a significant impact on the difference. The
more hostnames are associated with product-related categories before
the profiling commences, the better the quality of the generated
profiles.

\section{Related Work}
\label{sec::related}
\vspace{-0.1cm}
This work focuses on the development of a platform for the deployment
of deep learning methods in the communication network. Several
platform exist that can perform data analysis in a general streaming
setting; among these are four Apache projects \cite{carbone2015apache,
  storm2014storm, apacheKafka, sparkStreaming} and proprietary
solutions such as \cite{biem2010ibm}. However, they are not optimized
for deep learning-based analysis nor for dealing with network
data. The only solution optimized for the network setting (to the best
of our knowledge) is Blockmon~\cite{huici2012blockmon,
  simoncelli2013stream}; however, it does not support the use of deep
learning algorithms.

We can also find systems like Clipper \cite{crankshaw2016clipper} that
allow predictions in a streaming setting. However, they are not
optimized for working with high network data rates, nor do they
provide any help for the collection and preprocessing of the data.

Finally, several specific network problems have been solved using GPUs
in recent years~\cite{renart2015towards, han2010packetshader,
  wang2013wire} Our solution is general and can be applied to a wide
range of network problems ranging from traffic classification to user
profiling while leveraging acceleration on heterogeneous hardware such
as GPUs.

%
%
%
%
%
%
%
%
%
%
%
%
%
%
%
%
%

\vspace{-0.1cm}
\section{Conclusion and Future Work}
\label{sec::conclusion}
\vspace{-0.1cm}
We introduced \nettovec, a framework for data scientists to develop
and deploy machine learning methods in data communication
networks. The modular design of \nettovec provides enough flexibility
to develop completely different use cases in an efficient manner
without sacrificing performance. Our proof-of-concept evaluation shows
that \nettovec is able to efficiently profile network users at line
rate while presenting better efficiency than baseline methods.

As future work, we are looking into efficient methods for transferring
data into GPUs, as well as packet-based operations can be accelerated
by these devices. A separate, but complementary, goal is to enhance
\nettovec to make it seamless for data scientists to obtain network
data. Finally, we will add a new component that will allow the
platform to actually perform actions in order to apply real-time
re-configurations to the network.

\vspace{-0.1cm}


{\footnotesize \bibliographystyle{acm}
\bibliography{references}}

\end{document}